\documentclass[amsmath, amssymb, superscriptaddress, nofootinbib]{revtex4-2}

\usepackage{graphicx}
\usepackage{amsmath}
\usepackage{mathtools}
\usepackage{natbib}
\usepackage{caption}
\usepackage{float}
\usepackage{dcolumn}
\usepackage{bm}
\usepackage[colorlinks=true,citecolor=blue,linkcolor=blue,urlcolor=blue]
{hyperref}
\usepackage{subfig}

\begin{document}
	
	\title{Evolving wormhole formation in dRGT massive gravity}
	
	
	\author{Ayanendu Dutta}
	\email{ayanendudutta@gmail.com}
	\affiliation{Department of Physics, Jadavpur University, Kolkata-700032, India}
	
	\author{Dhritimalya Roy}
	\email{rdhritimalya@gmail.com}
	\affiliation{Department of Physics, Jadavpur University, Kolkata-700032, India}
	
	\author{Subenoy Chakraborty}
	\email{schakraborty.math@gmail.com}
	\affiliation{Department of Mathematics, Jadavpur University, Kolkata-700032, India}

	
\begin{abstract}
	In this study, we have examined the evolving wormhole solution within Einstein-massive gravity, considering traceless, barotropic, and anisotropic pressure fluids. We have conducted a comprehensive analysis of the constraints imposed by the constants derived from the wormhole solution. It is found that the wormhole throat, situated between two asymptotic universes, undergoes simultaneous expansion with acceleration. A detailed investigation of the energy conditions for traceless, barotropic, and anisotropic fluids suggests a wide range of possibilities for evolving wormhole configurations with non-exotic matter at the throat. The dependency of this feature on the various parameters arising from the study has also been examined.
\end{abstract}


\maketitle

\section{Introduction}
Wormholes serve as seamless connections between distinct universes or, at times, between remote regions within the same universe. The concept originated in 1935 with Einstein and Rosen \cite{einstein-rosen}, and the term `wormhole' was coined by Misner and Wheeler in 1957 \cite{misner1957}. The precise solutions to the Einstein Field Equation for a static, spherically symmetric `traversable' wormholes were successfully explored much later in 1988 by Morris and Thorne \cite{morris1988,morris1988_2}. Their investigation revealed that the energy-momentum component for such wormholes invariably violates the null energy condition \cite{morris1988, visser_lorentzian-wormhole}, the weakest classical energy condition, consequently violating all other energy conditions. Hence, the construction of traversable wormholes necessitates a substance with negative energy density, known as exotic matter. 

From the theoretical point of view, cosmology is the most promising field where the presence of exotic fluid is investigated the most. It is very much well-known that the acceleration of the universe is attributed to exotic matter characterized by the condition $\omega<-1/3$ and following the equation of state $p=\omega \rho$. Phantom energy, characterized by a parameter $\omega<-1$, exhibits distinct properties, including negative temperature and energy, and it contributes to a scenario known as the Big Rip, wherein its energy density evolves with the expansion of the universe \cite{Caldwell:2003vq}. Back in 1981, Sato and collaborators \cite{Sato:1981bf} explored the potential for dynamic wormhole formation during the inflationary era. Other facets of evolving wormholes at the Planck length scale have been examined by Roman \cite{Roman:1992xj} later. 
Subsequently, the cosmological principle asserts that at any given cosmic time, the universe displays to be homogeneous and isotropic over large scale structure, meaning it remains consistent regardless of shifts or rotations relative to any observer moving with the cosmic flow. This allows us to consider that it is invariant under special points and directions. The geometry that aligns with the cosmological principle is described by the Friedmann–Lemaitre–Robertson–Walker (FLRW) metric \cite{Trodden:2004st}. While the current status is accepted as undergoing an accelerated phase of expansion, the realistic tests of the structure formation e.g. sheets, filaments, dark halos etc. have challenged the homogeneity in various phases of epoch. However, the confirmation of isotropy still comes from observations of the cosmic microwave background radiation (CMBR). On the other hand, it is noted that on a small scale, we encounter self-gravitating solutions that persist irrespective of the accelerated expansion. Recent observations such as the galactic center black hole in M87 and Sagittarius A* in Event Horizon Telescope \cite{akiyama2019L1,akiyama2019L4,akiyama2022L12}, gravitational waves in LIGO/VIRGO collaboration \cite{abbott2016,abbott2017,abbott2019} have reportedly verified the validity of GR on a small scale.

Lemaitre and Tolman presented a realistic cosmological model, introducing the first framework allowing the study of inhomogeneous cosmology. Their model involves a spacetime filled with a perfect fluid characterized by the dust equation of state \cite{Lemaitre:1933gd,Tolman:1934za}. They developed geometries based on a spatially homogeneous, spherically symmetric background, later replaced by an inhomogeneous distribution at small scales. In essence, inhomogeneous cosmological models deviate from satisfying the cosmological principle but converge to the limit of the FLRW spacetime. Subsequently, numerous studies have delved into the exploration of inhomogeneous models, for example the Szekeres-Szafron model \cite{Szekeres:1975dx,collins1979,szafron1979,collins1979_2,Szafron:1977zza,Ferrando:2018bzh} is one of the models that enable the examination of spherically symmetric and inhomogeneous spacetimes, seamlessly transitioning to the cosmological background.

The time-dependent solutions for dynamic wormholes in inhomogeneous and spherically symmetric spacetime have been derived when considering a matter source with both radial and transverse stresses, as detailed in \cite{Bordbar:2010zz}. Various studies has uncovered time-dependent wormhole solutions as exact solutions on an inhomogeneous brane embedded in a 5-dimensional constant curvature bulk \cite{Heydari-Fard:2017oee}. The exploration of evolving Lorentzian wormholes, along with an analysis of the null energy condition (NEC) and weak energy condition (WEC), can be found in \cite{Kar:1994tz,Kar:1995ss,Hochberg:1998ii,Cataldo:2008pm,Cataldo:2008ku,Cataldo:2011zn,Cataldo:2013sma,KordZangeneh:2020jio,Cataldo:2012pw}, where some of the scenarios explored the feasibility of matters that satisfy the energy conditions. 

Notably, in modified gravity theories, the requirement for exotic matter is significantly reduced when constructing traversable wormholes. Numerous studies have delved into wormhole geometries and their associated energy conditions within various modified gravity frameworks, such as \cite{Golovnev:2018icm,Xavier:2024iwr,Kanzi:2020cyv,Ovgun:2018xys,Sakalli:2015taa,Sadeghi:2022sto,Sadeghi:2021pqg}. Since the introduction of the traversable wormhole model, there has been a keen interest in exploring the possibility of constructing wormholes using ordinary matter. Last year, a study has been conducted particularly on the behaviour of matter at the wormhole throat in Einstein gravity and in modified gravity theories which leads to the conclusion that wormholes can be constructed with ordinary matter in modified theories of gravity with specific constraints \cite{epl_paper}. There are numerous works that have conclusively obtained various static wormhole models using non-exotic matter \cite{fukutaka1989,hochberg1990,ghoroku1992,furey2005,bronnikov2010,kanti2011,kanti2012,harko2013,moraes2018}. 

Various models of modified gravity theory have been developed as extensions of Einstein's General Relativity to address cosmological phenomena. One significant candidate in this regard is the massive gravity theory. The detection of gravitational waves from the mergers of black holes and massive stars by LIGO and VIRGO \cite{abbott2016,abbott2017,abbott2019} has imposed stringent constraints on the mass of the graviton, intensifying interest in the theoretical aspects of massive gravity theory. The de Rham-Gabadadze-Tolley (dRGT) massive gravity \cite{derham2010,derham2011}, presented as a nonlinear generalization to overcome the van Dam-Veltman-Zakharov discontinuity \cite{dam1970,zakharov1970} of Fierz-Pauli linear massive gravity action \cite{fierz1939}, has gained attention. The dRGT theory is particularly promising as it addresses the long-standing issue of the Boulware-Deser ghost instability \cite{boulware1972}, a problem often associated with the introduction of nonlinear generalizations in massive gravity theory. The improvement of Lagrangian of gravity with mass and polynomial interaction terms, up to the fifth order in non-linearities, lead to the avoidance of ghost-like pathologies in the four-dimensional covariant nonlinear massive gravity theory. The effective field theory is then consistent for the Boulware-Deser issue in the decoupling limit of all order. They showed that the linear and some nonlinear mixing terms between helicity-0 and helicity-2 modes can be absorbed through a local variable transformation, which naturally leads to the cubic, quartic, and quintic Galileon interactions that were originally introduced in a different context. It is also noted that, in the decoupling limit, the mixing between these helicity modes can extend only up to quartic order in the decoupling limit.

In the dRGT theory, the additional parameters generated by mass of the gravitons may allow the theory to address both the dark matter and dark energy problems in the galactic and extragalactic scenarios within a unified framework. Consequently, the rotation curve of galaxies is influenced by the generation of a dark matter halo, and massive gravitons can act as the dark matter halo, resulting in asymptotically flat rotation curves. In a study by Panpanich \textit{et al.} \cite{panpanich2018}, the dRGT model was successfully fitted to observational data for Milky Way rotation curves and Low Surface Brightness (LSB) galaxies without the inclusion of additional dark matter. The consistency of the dRGT model with the Navarro-Frenk-White (NFW) profile was also noted in this context. The massive gravity modification is also allowed for the explanation of late-time cosmic acceleration; however, it does not provide a justification for past inflation. In a revised framework, it has been established that the massive gravity theory aligns with the Plank 2018 data and remains consistent when combined with BK18 and BAO, as indicated in \cite{afshar2022, bk18}.

The thermodynamics of dRGT theory has been investigated in ref. \cite{Beigmohammadi:2023ves}. They examined the generalized second law of thermodynamics (GSLT) in massive gravity framework. So they analyzed whether the time variation of matter entropy as well as horizon entropy is an increasing function of time. They have considered a FLRW universe with pressureless matter and bounded by the apparent horizon. At first they studied the cosmological background for the dRGT on de Sitter models and then examined the GSLT with different model parameters. The results of the above study can be written pointwise as: (i) the fractional deviation of the Hubble parameter (i.e. $ \delta H/H $) from $ \Lambda $CDM on de Sitter model are of the order of $ 10^{-3} $ (i.e. $ \mathcal{O}(10^{-3}) $), (ii) the equation of state parameter behaves like phantom fluid (i.e. $ \omega_{DE} <-1 $), (iii) the entropy of matter violates the second law of thermodynamics i.e. $ T_A \dot{S}_m <0 $, but adding with horizon entropy, the total entropy satisfies the GSLT on the apparent horizon. On the other hand, the phenomenology within the framework of dRGT massive gravity for compact objects holds significant promise and has garnered considerable interest. Numerous studies have delved into black holes, black strings, rotating black string solutions, stability, and the greybody factor for charged black holes and black strings in the dRGT model \cite{Nieuwenhuizen:2011sq,Burikham:2017gdm,Tannukij:2017jtn,Boonserm:2017qcq,Ponglertsakul:2018smo,Boonserm:2019mon,Ghosh:2019eoo} along with their thermodynamical properties \cite{Ghosh:2015cva,Ma:2020jls,Chunaksorn:2022whl,Chen:2023pgs}. These investigations differ notably from conventional studies due to the presence of massive gravitons.
The nonlinear interaction of gravitons generates density and pressure, behaving akin to dark energy. Consequently, the energy-momentum tensor of massive gravitons inherently exhibits a violation of energy conditions \cite{sushkov2015}. Thus, the presence of a wormhole in massive gravity can be a natural outcome, and exploring this particular object holds additional significance. One may check the recent investigation of wormhole solutions in dRGT massive gravity as given by \cite{Dutta:2023wfg,tangphati2020,kamma2021} for instance. An interesting feature coming out of \cite{Dutta:2023wfg} is the presence of repulsive gravity effect in the theory produced by the strong repulsion of massive gravitons, and this may lead to the violation of asymptotic flatness. This study is also emerged as a noteworthy candidate for static traversable wormholes with non-exotic matter at the throat, thereby enhancing the possibility of non-exotic matter wormholes in this theory.

Therefore, in this article, we will examine the  time-dependent evolving wormhole solution within the framework of Einstein-massive gravity. The study starts with the field equations of dRGT massive gravity, the discussion of spacetime geometry, and field equations of the metric in section \ref{field_eq}. The evolving wormhole solutions respectively in traceless fluid, barotropic equation of state (EOS), and anisotropic pressure fluid are computed in section \ref{solution}. Section \ref{energy-condition} is dedicated for a thorough examination of the energy conditions. Finally, the study ends with a conclusive discussion in section \ref{discussions}.

We employ natural units consistently throughout the study, where $ G=c=1 $.

\section{The field equations}\label{field_eq}
The de Rham-Gabadadze-Tolley (dRGT) massive gravity can be represented as Einstein gravity interacting with a scalar field. Consequently, its action comprises the familiar Einstein-Hilbert action combined with appropriate nonlinear interaction terms, which are defined as follows \cite{derham2010,derham2011}
\begin{equation}\label{eq1}
	S = \int d^4 x \sqrt{-g} \bigg(\frac{1}{16\pi}\Big[ R + m^2_g \mathcal{U}(g,\phi^a) \Big] +\mathcal{L}_{m} \bigg),
\end{equation}
where the matter Lagrangian is $ \mathcal{L}_{m} $, and the determinant $ g $ corresponds to the metric tensor $ g_{\mu \nu} $. The self-interacting potential for graviton $ \mathcal{U} $ modifies the usual gravitational sector with the graviton mass $ m_g $. In four-dimension, it can be written as
\begin{eqnarray}\label{eq2}
	\mathcal{U} = \mathcal{U}_2 + \alpha_3 \mathcal{U}_3 + \alpha_4 \mathcal{U}_4 ,
\end{eqnarray} 
where $ \alpha_3 $ and $ \alpha_4 $ are two dimensionless free parameters of the massive gravity theory. The functional forms of the potential i.e., $ \mathcal{U}_j $ can be expressed in terms of the metric $ g $ and St\"uckelberg scalar $ \phi^a $ as
\begin{align}\label{eq3}
	\mathcal{U}_2 &= [\mathcal{K}]^2 - [\mathcal{K}^2], \nonumber \\
	\mathcal{U}_3 &= [\mathcal{K}]^3 - 3[\mathcal{K}][\mathcal{K}^2] + 2 [\mathcal{K}^3], \nonumber \\
	\mathcal{U}_4 &= [\mathcal{K}]^4 - 6[\mathcal{K}]^2[\mathcal{K}^2] + 8[\mathcal{K}][\mathcal{K}^3] + 3[\mathcal{K}^2]^2 -     6[\mathcal{K}^4], \nonumber \\
	\mathcal{{K}^{\mu}}_{\nu} &= \delta^{\mu}_{\nu} - \sqrt{g^{\mu\lambda} \partial_{\lambda}\phi^a \partial_{\nu}\phi^b \mathcal{F}_{ab} }.
\end{align}
In these equations, $ [\mathcal{K}] $ denotes the trace of $ {K}^{\mu}_{\nu} $, where $ (\mathcal{K}^i)^{\mu}_{\nu} = \mathcal{K}^\mu_{\rho_1} \mathcal{K}^{\rho_1}_{\rho_2} ...\mathcal{K}^{\rho_i}_\nu $. The interaction terms can be identified as symmetric polynomials of $ \mathcal{K} $. For a specific order, the coefficients for each combination are selected to ensure that these terms do not introduce higher derivative terms in the equations of motion. Notably, this definition of $ \mathcal{K} $ is not exclusive, as an alternating action can be achieved with a different definition of $ \mathcal{K} $. 

The four scalar St\"uckelberg field $ \phi^a $ is introduced in the theory to restore general covariance. This field is analogous to the reference fiducial metric
\begin{equation}\label{eq4}
	\mathcal{F}_{ab} = diag(0,0,c^2,c^2 \sin^2 \theta) ,
\end{equation}
where $ c $ is a positive constant having the dimension of length. The reference metric's dependence solely on the spatial components ensures that general covariance is maintained in the t and r coordinates, but it is disrupted in the two spatial dimensions. One can also consider a more comprehensive reference metric that does not maintain diffeomorphism invariance in the r-direction. For example, to preserve rotational symmetry on the sphere and general time reparametrization invariance, a natural choice could be \( \mathcal{F}_{ab} = \text{diag}(0, 1, c^2, c^2 \sin^2 \theta) \). Another way to break diffeomorphism invariance in the r-direction could involve a different form of \( \mathcal{F}_{ab} \), where \( \sin^2 \theta \mathcal{F}_{\theta\theta} = \mathcal{F}_{\phi\phi} = F(r) \), with all other components set to zero. This implies that exploring various forms for the reference metric can lead to a range of new solutions, making massive gravity with this reference metric a compelling subject for researchers. However, this paper does not aim to delve into such investigations.

Note that the unitary gauge is realized as, $ \phi^a = x^\mu \delta^a_\mu $ \cite{Vegh:2013sk}, such that
\begin{eqnarray}\label{eq5}
	\sqrt{g^{\mu\lambda} \partial_{\lambda}\phi^a \partial_{\nu}\phi^b \mathcal{F}_{ab} }= \sqrt{g^{\mu \lambda} \mathcal{F}_{\lambda \nu} }.
\end{eqnarray}
In the above gauge, the tensor \( g_{\mu\nu} \) represents the observable metric corresponding to the five degrees of freedom of the massive graviton. It is important to note that since the Stückelberg scalars transform according to coordinate transformations, fixing these scalars—such as by selecting the unitary gauge—means that any subsequent coordinate transformation will violate the gauge condition and cause further modifications to the Stückelberg scalars.

To proceed, the effective energy-momentum tensor $ X_{\mu \nu} $ of the massive gravitons is established as
\begin{widetext}
	\begin{eqnarray}\label{eq11}
		\nonumber	X^\mu_\nu &=& \mathcal{K}^\mu_\nu -[\mathcal{K}] \delta^\mu_\nu -\alpha \left[ (\mathcal{K}^2)^\mu_\nu -[\mathcal{K}]\mathcal{K}^\mu_\nu +\frac12 \delta^\mu_\nu \left( [\mathcal{K}]^2 -[\mathcal{K}^2] \right) \right] \\
		&&+3\beta \left[ (\mathcal{K}^3)^\mu_\nu -[\mathcal{K}](\mathcal{K}^2)^\mu_\nu +\frac12 \mathcal{K}^\mu_\nu \left( [\mathcal{K}]^2- [\mathcal{K}^2] \right) \right] -3\beta \left[ \frac16 \delta^\mu_\nu \left( [\mathcal{K}]^3- 3[\mathcal{K}] [\mathcal{K}^2]+ 2 [\mathcal{K}^3] \right) \right],
	\end{eqnarray}
\end{widetext}
where $ \alpha $ and $ \beta $ are two new dimensionless arbitrary constants used here to accommodate the parameters $ \alpha_3 $ and $ \alpha_4 $, as given by
\begin{eqnarray}\label{eq12}
	\alpha = 1 + 3\alpha_3\,,\qquad \beta = \alpha_3 + 4\alpha_4.
\end{eqnarray}
Eq. \eqref{eq11} obeys the conservation relation by virtue of the Bianchi identities as $ \nabla^\mu X_{\mu \nu}=0 $, where $ \nabla^\mu $ represents the covariant derivative, compatible with the metric $ g_{\mu \nu} $. Therefore, by definition, the principle pressure form of the energy momentum tensor of massive gravitons is given by
\begin{equation}\label{ener-mom_massive}
	\frac{m^2_g}{8\pi}X_{\mu \nu}= -(\rho^{(g)}+p_t^{(g)})u_\mu u_\nu -p_t^{(g)} g_{\mu \nu} -(p_r^{(g)}-p_t^{(g)})\chi_\mu \chi_\nu,
\end{equation}
where $ u^\mu $ is the timelike four-vector and $ \chi^\mu $ is the spacelike vector orthogonal to the timelike unit vector, satisfying $ u_\mu u^\mu=-1 $ and $ \chi_\mu \chi^\mu=1 $. $ \rho^{(g)},p_r^{(g)} $ and $ p_t^{(g)} $ respectively define the total energy density, radial pressure and transverse pressure for the massive gravitons.

Hence, we are now ready to vary the action with respect to metric $ g_{\mu \nu} $ and obtain the desired field equation for dRGT-Einstein massive gravity as given by
\begin{equation}\label{eq6}
	G_{\mu \nu}=  8\pi T_{\mu \nu}- m_g^2 X_{\mu \nu}.
\end{equation}

The variation of the trace of the matter field's energy-momentum tensor, represented by $ T= g^{\mu \nu} T_{\mu \nu} $, can be expressed as
\begin{equation}\label{eq7}
	\frac{\delta(g^{\alpha \beta} T_{\alpha \beta})}{\delta g^{\mu \nu}} =T_{\mu \nu} +\Theta_{\mu \nu},
\end{equation}
where $ \Theta_{\mu \nu} $ and $ T_{\mu \nu} $ are given by
\begin{eqnarray}\label{eq8}
	\Theta_{\mu \nu} &\equiv& g^{\alpha \beta} \frac{\delta T_{\alpha \beta}}{\delta g^{\mu \nu}},
	\label{eq8}\\
	T_{\mu \nu} &\equiv& g_{\mu \nu} \mathcal{L}_m - \frac{2 \partial(\mathcal{L}_m)}{\partial g^{\mu \nu}}.
	\label{eq9}
\end{eqnarray}
\\
Given the standard Lagrangian matter density as $ \mathcal{L}_m=\rho $, the tensor $ \Theta $ can be expressed as $ \Theta_{\mu \nu}= -2T_{\mu \nu}+\rho g_{\mu \nu} $. On the other hand, assuming a timelike unit vector $ u_\mu $, and the corresponding spacelike unit vector $ \chi_\mu $ which is orthogonal to timelike unit vector, such that $ \chi_\mu \chi^\mu=1 $ and $ u_\mu u^\mu=-1 $, one can readily write down the $ T_{\mu \nu} $ in the principal pressure form as
\begin{equation}\label{EM-einstein}
	T_{\mu \nu}= (\rho+p_t)u_\mu u_\nu +p_t g_{\mu \nu} +(p_r-p_t)\chi_\mu \chi_\nu,
\end{equation}

Note that, in the field equation, $ T_{\mu \nu} $ represents the energy-momentum tensor of the Einstein gravity sector whereas, for the massive gravity sector, the same is written in Eq. \eqref{ener-mom_massive}. Hence, one may interpret the coupling of whole matter field as the exchange of energy and momentum between usual matter (sourced by Einstein gravity) and the massive gravitons, expressed as
\begin{equation}\label{EM-total}
	T^{tot}_{\mu \nu}= diag\left( -\rho-\rho^{(g)}, p_r+p_r^{(g)}, p_t+p_t^{(g)}, p_t+p_t^{(g)} \right).
\end{equation}

Now, to solve the field equations, one may consider the metric ansatz for the time-dependent dynamic wormhole as \cite{Roman:1992xj}
\begin{equation}\label{metric-dynamic}
	ds^2= -e^{2 \Phi(r)}dt^2+ a(t)^2 \left[ \frac{dr^2}{1- \frac{b(r)}{r}} +r^2 d\Omega^2 \right] ,
\end{equation}
where $ \Phi(r) $, $ b(r) $ and $ a(t) $ are respectively the redshift function, shape function and cosmic scale factor, and $ d\Omega^2= d\theta^2 + \text{sin}^2 \theta d\phi^2 $. With all these ingredients in hand, it is straightforward to compute the density and pressure components $ \rho^{(g)}(r,t) $ and $ p^{(g)}_{r,\perp}(r,t) $, for the massive gravitons. From equations \eqref{eq3}, \eqref{eq11}, \eqref{ener-mom_massive} and \eqref{metric-dynamic}, we obtain
\begin{eqnarray}
	\rho^{(g)}(r,t) &\equiv& \frac{m_g^2}{8\pi}{X^t}_t 
	= \frac{1}{8\pi}\left(\Lambda - \frac{2 \gamma}{ar} - \frac{\epsilon}{a^2 r^2} \right),
	\label{eq14}\\
	p_r^{(g)}(r,t) &\equiv& -\frac{m_g^2}{8\pi}{X^r}_r = -\frac{1}{8\pi}\left(\Lambda - \frac{2 \gamma}{ar} - \frac{\epsilon}{a^2 r^2} \right),
	\label{eq15}\\
	p_{\theta, \phi}^{(g)}(r,t) &\equiv& -\frac{m_g^2}{8\pi}X_{\theta,\phi}^{\theta,\phi }
	= -\frac{1}{8\pi}\left(\Lambda - \frac{\gamma}{ar} \right).
	\label{eq16}
\end{eqnarray}
where the effective cosmological constant $ \Lambda $ and two new parameters $ \gamma $ and $ \epsilon $ are introduced. They are expressed as a linear combination of the parameters 
$ \alpha $ and $ \beta $, and are given by
\begin{eqnarray}
	\Lambda \equiv -3m_g^2(1+ \alpha + \beta), \qquad \gamma \equiv -m_g^2c(1 + 2\alpha + 3\beta),
	\qquad \epsilon \equiv m_g^2c^2(\alpha + 3\beta).
	\label{massive-parameters}
\end{eqnarray}
Therefore, by setting $ m_g=0 $, we recover the usual solutions in Einstein's GR. For \( c = 0 \), where \(\gamma = \epsilon = 0\), the solution can be categorized based on the values of \(\alpha\) and \(\beta\). If \(1 + \alpha + \beta < 0\), it results in the de Sitter solution. Conversely, if \(1 + \alpha + \beta > 0\), it leads to the Anti-de Sitter solution. We will see in Section \ref{solution}, that the term \(\gamma a/ r\) is a characteristic term of this solution, which distinguishes it from other solutions. However, it also creates unavoidable difficulty in the evolving wormhole solution. The constant potential \(\epsilon\) corresponds to the global monopole term, which arises naturally from the graviton mass. Typically, a global monopole solution stems from a topological defect in high-energy physics during the early universe, resulting from gauge-symmetry breaking \cite{Huang:2014oga,Tamaki:2003kv}. It is possible to choose different combinations of $ \alpha $ and $ \beta $, that may lead to different solutions. To avoid complexities present in the theory, sometimes particular relations of these parameters are chosen such that $ \gamma $ or $ \epsilon $ from Eq. \eqref{massive-parameters} can be neglected. For instance, $ \alpha=-3\beta $ neglects $ \epsilon $ \cite{Berezhiani:2011mt}. Furthermore, $ \beta=\alpha^2/3 $ yields Schwarzschild–de Sitter solution \cite{Kodama:2013rea}.

For the discussion on the evolving wormhole geometry, the basic criteria for ensuring the traversability of the wormhole model \eqref{metric-dynamic} entail the following Morris-Thorne traversability conditions:
\begin{enumerate}
	\item The construction of the wormhole involves gluing two asymptotically flat regions through a throat. The throat radius is determined by a global minimum, denoted as $ r=r_0 $ implying $ b(r_0)=r_0 $. Consequently, the radial coordinate spans the interval $ r \in [\ r_0,\infty )\ $.
	\item For the avoidance of horizons and singularities, it is imperative that the redshift function $ \Phi(r) $ remains finite across all points, ensuring $ e^{\Phi(r)}>0 $ for all $ r>r_0 $. Within this framework, the ultrastatic wormhole holds particular significance as it defines the zero-tidal-force wormhole, where $ \Phi(r)=0 $, thereby implying that $ e^{\Phi(r)}=1 $. In particular, in a gravitational acceleration-free frame, a particle released from rest remains stationary \cite{morris1988,cataldo2017}.
	\item The flaring-out condition $ -rb'(r)+b(r)>0 $ must be satisfied at or in close proximity to the throat at $ r=r_0 $.
	\item Finally, the asymptotic flatness requires $ \Phi(r)\rightarrow 0 $ and $ b(r)/r \rightarrow 0 $ as $ r \rightarrow \infty $.
\end{enumerate}
Readers are referred to \cite{morris1988,tello2021}, for detailed discussions on the requirement of traversability in Morris-Thorne type wormholes.

Note that, in the orthonormal frame, the metric signature is denoted by $ g_{\hat{\mu}\hat{\nu}}=\text{diag}(-1,1,1,1) $, and the corresponding vectors are
\begin{eqnarray}
	\textbf{e}_{\hat{0}} &=& e^{-\Phi} \textbf{e}_t, \nonumber \\
	\textbf{e}_{\hat{1}} &=& \frac{\textbf{e}_r \sqrt{1-b/r}}{a}, \nonumber \\
	\textbf{e}_{\hat{2}} &=& \frac{\textbf{e}_\theta}{a r}, \\
	\textbf{e}_{\hat{3}} &=& \frac{\textbf{e}_\phi}{a r \text{sin}\theta}. \nonumber
\end{eqnarray}
Finally, we are now able to compute the Einstein tensor components for the spacetime \eqref{metric-dynamic} as given by
\begin{eqnarray}
	G_{00} &=& \frac{b'}{r^2 a^2} -3H^2 e^{-2\Phi},
	\label{g00}\\
	G_{11} &=& -\frac{b}{r^3 a^2}+ \frac{2\Phi'}{r a^2} \left( 1- \frac{b}{r} \right) -e^{-2\Phi}\left(2\dot{H}+3H^2\right),
	\label{g11}\\
	G_{22} &=& G_{33} =	\frac{1}{a^2}\left( 1- \frac{b}{r} \right) \left(\Phi''+\Phi'^2\right) +\frac{1}{2r^3 a^2}\left(-rb'+b\right) -\frac{\Phi'}{2r^2 a^2}\left(rb'+b-2r\right) -e^{-2\Phi}\left(2\dot{H}+3H^2\right).
	\label{g22}
\end{eqnarray}
where the prime and overdot respectively defines differentiation with respect to $ r $ and $ t $, and $ H(a)=\dot{a}(t)/a(t) $. Notice that the EM components of Einstein gravity as given by Eq. \eqref{EM-einstein} satisfies the energy conservation law such that $ \nabla^\mu T_{\mu \nu}=0 $. Consequently, one can compute the total EM sector defined by the coupled perfect fluid-massive graviton system of Eq. \eqref{EM-total} also obey $ \nabla^\mu T^{tot}_{\mu \nu}=0 $. Therefore, referring to Eq. \eqref{EM-einstein} and Einstein tensor components, the field equations take the form
\begin{eqnarray}
	\rho &=& \frac{b'}{8 \pi r^2 a^2}+ \frac{3H^2}{8\pi} e^{-2\Phi} - \frac{1}{8\pi}\left(\Lambda - \frac{2 \gamma}{ar} - \frac{\epsilon}{a^2 r^2} \right),
	\label{field-1} \\
	p_r &=& -\frac{b}{8 \pi r^3 a^2}+\left( 1-\frac{b}{r} \right) \frac{\Phi'}{4\pi r a^2} -\frac{e^{-2\Phi}}{8\pi} \left(2\dot{H}+3H^2\right) + \frac{1}{8\pi}\left(\Lambda - \frac{2 \gamma}{ar} - \frac{\epsilon}{a^2 r^2} \right),
	\label{field-2} \\
	p_t &=& \frac{1}{8\pi a^2} \left( 1-\frac{b}{r} \right) \left(\Phi'' + \Phi'^2 \right)+ \left(\frac{-r b'+b}{16\pi r^3 a^2}\right) -\left(\frac{rb'+b-2r}{16\pi r^2 a^2}\right) \Phi' -\frac{e^{-2\Phi}}{8\pi} \left(2\dot{H}+3H^2\right) + \frac{1}{8\pi}\left(\Lambda - \frac{\gamma}{ar} \right).
	\label{field-3}
\end{eqnarray}

Now, to construct the dynamic wormhole solution, the simplified method considers restricted choices of $ \Phi(r), ~b(r) $ and $ a(t) $ to analyze the corresponding dynamics. However, this may neglect the dependence of the solutions on massive gravity parameters. The alternative method considers specific constraints in the energy density, radial and transverse pressures such as isotropic, anisotropic or the traceless fluids. In this study, we consider various fluid solutions for a particular class of wormhole i.e. the ultrastatic wormhole for which the field equations are modified to
\begin{eqnarray}
	\rho &=& \frac{b'}{8 \pi r^2 a^2}+ \frac{3H^2}{8\pi}  - \frac{1}{8\pi}\left(\Lambda - \frac{2 \gamma}{ar} - \frac{\epsilon}{a^2 r^2} \right),
	\label{field-eq1} \\
	p_r &=& -\frac{b}{8 \pi r^3 a^2} -\frac{2\dot{H}+3H^2}{8\pi} + \frac{1}{8\pi}\left(\Lambda - \frac{2 \gamma}{ar} - \frac{\epsilon}{a^2 r^2} \right),
	\label{field-eq2} \\
	p_t &=& \left(\frac{-r b'+b}{16\pi r^3 a^2}\right) -\frac{2\dot{H}+3H^2}{8\pi} + \frac{1}{8\pi}\left(\Lambda - \frac{\gamma}{ar} \right).
	\label{field-eq3}
\end{eqnarray}
Thus, we are all set to delve into the wormhole solutions and discuss the dynamics.

\section{Wormhole Solution}\label{solution}
In this section, we consider three particular choices of the pressure components to compute the solutions, i.e., (i) Traceless fluid, (ii) Barotropic EOS (Equation of state), and (iii) Anisotropic fluid. Notice that, from the field equations \eqref{field-eq1}, \eqref{field-eq2}, \eqref{field-eq3}, one can identify that the isotropic fluid ($ p_t=p_r $) solution has some particular limitations as it does not incorporate the cosmic scale factor $ a(t) $. Hence, it is kept aside, however the limitations are briefly discussed in \textit{Discussions} (Sec. \ref{discussions}).

In the cosmological scenarios, studying traceless fluids, barotropic fluids, and anisotropic fluids is important for several reasons, especially when considering the evolution of the universe in the framework of general relativity and modern cosmology \cite{Peebles:1994xt,Misner:1973prb,Weinberg2008,Planck:2018vyg,Copeland:2006wr}. Traceless fluids are often associated with radiation-dominated phases in the universe. The stress-energy tensor for radiation is traceless because the pressure is one-third of the energy density. These fluids are critical in early cosmological models, particularly in the context of the radiation-dominated era, which occurred shortly after the Big Bang.

Subsequently, Barotropic fluids are defined by a relationship between the pressure \( p \) and energy density \( \rho \), usually written as \( p = \omega \rho \), where \( \omega \) is a constant. The value of \( \omega \) dictates the cosmological behavior of different eras in the universe. For example (a) \( \omega = -1 \) represents \(\Lambda\)CDM, (b) \( \omega < -1 \) represents phantom field, and (c) \( -1 < \omega < -1/3 \) describes quintessence. Studying barotropic fluids is key in exploring the transition between different cosmological phases and understanding the evolution of the universe’s expansion rate. Specifically, dark energy and the accelerated expansion of the universe are modeled using a barotropic equation of state with \( \omega \approx -1 \).

On the other hand, anisotropic fluids are important in studying the early universe and certain cosmological solutions, such as Bianchi models, which describe universes with anisotropic spacetime metrics. Anisotropic fluid models can also provide insights into the evolution of the primordial gravitational waves and the CMB anisotropies. The presence of anisotropic fluids can significantly alter the dynamics of spacetime, leading to unique solutions in Einstein’s field equations that might describe exotic cosmological scenarios.

\subsection{Traceless fluid ($ -\rho+p_r+2p_t=0 $)}
For the choice of traceless matter fluid, the trace of EM tensor $ T=0 $ so that $ -\rho+p_r+2p_t=0 $. The condition provide the following expression:
\begin{eqnarray}
	\frac{b'(r)}{r^2}+3a(t) \ddot{a}(t)+3\dot{a}(t)^2 -2\Lambda a(t)^2 + \frac{\epsilon}{r^2} +\frac{3\gamma a(t)}{r}=0.
	\label{traceless-master}
\end{eqnarray}
where \(\Lambda\) and \(\epsilon\) represent the effective cosmological constant and the global monopole term, respectively. Additionally, \(\gamma\) and \(\epsilon\) together characterize the deviation from the Schwarzschild de Sitter and Anti-de Sitter solutions. In black hole solutions within the framework of massive gravity, \(\gamma < 0\) behaves similarly to black holes surrounded by a quintessence field when the \(\Lambda\) and \(\epsilon\) terms are absent. Conversely, without \(\gamma\), black holes are found to conform to the D bound and the Bekenstein bound \cite{Hadi:2023wkn}. However, in wormhole configurations, \(\gamma\) does not exhibit any significant physical effect. It is important to note that the term involving \(\gamma\) in \eqref{traceless-master} presents a major challenge in the process of variable separation. Therefore, the only feasible approach is to eliminate \(\gamma\) by imposing constraints on \(\alpha\) and \(\beta\). This adjustment is not particularly problematic, as explained in the previous section with appropriate references.

By considering \(\alpha = -(1 + 3\beta)/2\), which sets \(\gamma = 0\), Eq. \eqref{traceless-master} can be separated with a constant \(c_1\) as follows:
\begin{eqnarray}
	\frac{b'(r)}{r^2} + \frac{\epsilon}{r^2} = c_1,
	\label{traceless-separation1} \\
	3a(t) \ddot{a}(t)+3\dot{a}(t)^2 -2\Lambda a(t)^2 = -c_1.
	\label{traceless-separation2}
\end{eqnarray}
where $ c_1 $ is the separation constant. Now, the shape function $ b(r) $ can be easily solved by imposing the throat condition $ b(r=r_0)=r_0 $ and is given by
\begin{eqnarray}
	b(r)=r_0+\frac{c_1}{3}\left(r^3-r_0^3\right) - \epsilon(r-r_0),
	\label{traceless-shape}
\end{eqnarray}
and the scale factor as
\begin{eqnarray}
	a(t)^2=\frac{c_1}{2\Lambda}+A_1 e^{2\sqrt{\Lambda/3}t}+B_1 e^{-2\sqrt{\Lambda/3}t},
	\label{traceless-scale}
\end{eqnarray}
where $ A_1 $ and $ B_1 $ are the arbitrary integration constants. It is interesting to note that for the choice $ B_1=0 $, the universe evolves from an emergent scenario in the infinite past i.e. $ a(t)\rightarrow a_0 (=\sqrt{c_1/2\Lambda}) $, $ H\rightarrow 0 $ as $ t\rightarrow -\infty $. Thus the above wormhole configuration evolves from an emergent phase. Now due to dependence of $ \Lambda $ on the mass of the massive gravitons, it is reasonable to think that the present massive gravity theory may avoid the big-bang singularity.

Now, for the validity of the flaring-out condition for the present wormhole configuration it is found that the radial coordinate `$ r $' is restricted as
\begin{eqnarray}
	r < r_0 \left[\frac12 \left(\frac{3(1+\epsilon)}{c_1 r_0^2}-1\right)\right]^{1/3}=r_1 \text{(say)}.
	\label{flare-traceless}
\end{eqnarray}
Now, the above restriction on $ r $ puts two restrictions on the throat radius $ r_0 $, namely
\\ \\
(i) $ \frac{3(1+\epsilon)}{c_1 r_0^2}-1>0 \qquad \text{i.e.} ~r_0^2<\frac{3(1+\epsilon)}{c_1} $,
\\
(ii) $ \frac12 \left(\frac{3(1+\epsilon)}{c_1 r_0^2}-1 \right) >1 \qquad \text{i.e.} ~r_0^2<\frac{(1+\epsilon)}{c_1} $,
\\
i.e.
\begin{equation}
	r_0^2<\frac{(1+\epsilon)}{c_1}.
	\label{flare-traceless2}
\end{equation}
Thus choosing the separation constant $ c_1 $ to be large +ve value, it is possible to make the throat radius arbitrarily small. Further, the present wormhole configuration is restricted to a finite region: $ r_0 \le r < r_1 $. Thus, this finite wormhole evolves from an initial emergent scenario.

\subsection{Barotropic EOS ($ p_r=\omega \rho $)}
\begin{figure*}[h!]
	\centering
	\includegraphics[height=5cm]{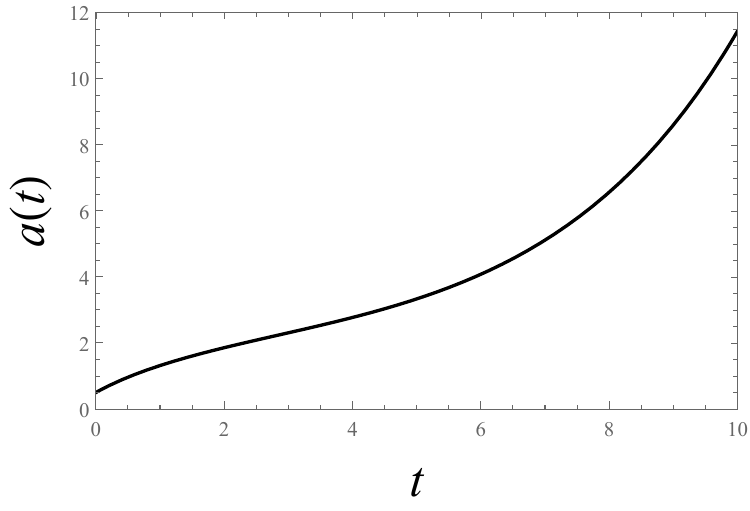}
	\caption{Plots showing the behaviour of scale factor $ a(t) $ of Eq. \eqref{eos-scale1} with cosmic time $ t $ for $\omega=-\frac13,~\Lambda=0.3,~c_2=1$, and $ a_0=0.5 $.}
	\label{fig:eos-scale1}
\end{figure*}

For the Equation of state (EOS) in barotropic matter the radial pressure and energy density are related by $ p_r=\omega \rho $, where $ \omega $ be the equation of state parameter. So, substituting Eq. \eqref{field-eq1} and \eqref{field-eq2} into this equation, we have
\begin{eqnarray}
	\frac{r \omega b'(r)+b(r)}{r^3}+2a(t) \ddot{a}(t)+(1+3\omega)\dot{a}(t)^2 -(1+\omega) \Lambda a(t)^2 +(1+\omega)\frac{\epsilon}{r^2} +2(1+\omega) \frac{\gamma a(t)}{r}=0.
	\label{eos-master}
\end{eqnarray}
Similar to the traceless fluid, this equation is also non-separable, and interestingly for the same approximation i.e. $ \gamma=0 $, it can be separated into $ r $ and $ t $ dependent functions as
\begin{eqnarray}
	\frac{r \omega b'(r)+b(r)}{r^3} +(1+\omega)\frac{\epsilon}{r^2} &=& c_2, 
	\label{eos-separation1} \\
	2a(t) \ddot{a}(t)+(1+3\omega)\dot{a}(t)^2 -(1+\omega) \Lambda a(t)^2 &=& -c_2,
	\label{eos-separation2}
\end{eqnarray}
with $ c_2 $, the separation constant. Now, solving the first order differential equation \eqref{eos-separation1}, the throat radius is given by
\begin{eqnarray}
	b(r) = 
	\begin{cases}
		r_0 \left(\frac{r}{r_0}\right)^{3} -\epsilon \left[r-r_0 \left(\frac{r}{r_0}\right)^3\right] -3 c_2 r^3 \ln\left(\frac{r}{r_0}\right), & \text{for}~ \omega=-\frac13 .
		\\
		r_0\left(\frac{r_0}{r}\right)^{1/\omega} -\epsilon \left[r-r_0 \left(\frac{r_0}{r}\right)^{1/\omega}\right] +\frac{c_2}{1+3\omega} \left[r^{3}- r_0^3 \left(\frac{r_0}{r}\right)^{1/\omega} \right], & \text{for}~ \omega \ne -\frac13 .
	\end{cases} 
	\label{eos-shape}
\end{eqnarray}
The scale factor $ a(t) $ can be obtained from the second order non-linear differential equation \eqref{eos-separation2} as
\begin{eqnarray}
	a(t)=
	\begin{cases}
		\sqrt{\frac{3A_2}{\Lambda}-\frac{9c_2^2}{16 \Lambda^2}} ~  \text{Sinh}\left[2 \sqrt{\frac{\Lambda}{3}} (t-t_0)\right] +\frac{3c_2}{4\Lambda}, & ~\text{for}~ \omega=\frac13.\\
		\sqrt{\frac{3c_2}{2(\Lambda+3 B_2)}} ~ \text{Sinh}\left[ \sqrt{\frac{\Lambda}{3}+B_2} \left(t-t_0 \right)\right], & ~ \text{for}~ \omega=-1.
	\end{cases}
\label{eos-scale2}
\end{eqnarray}
where $ t_0,~A_2,~B_2 $ are integration constants. Now, for $ \omega=-\frac13 $, $ t $ can be obtained as an integral form of the scale factor as
\begin{eqnarray}
	(t-t_0)= \int \frac{da}{\sqrt{\frac{\Lambda}{3}a^2 -c_2 \ln(a)+a_0}},
	\label{eos-scale1}
\end{eqnarray}
and the variation of this scale factor against $ t $ has been plotted in Fig. \ref{fig:eos-scale1}. For flaring-out condition to be satisfied, the restriction on the throat radius can be presented in the following tabular form (See table \ref{details-barotropic_table}).

\begin{table*}[!]
	\centering
	\begin{tabular}{ ||p {5 cm}| p {12 cm}|| }
		\hline
		~~\textbf{Case I}			&	\textbf{$ \omega=-\frac13 $} \\
		
		\hline
		~~Flaring-out relation	& 
		\begin{eqnarray}
			r > r_0 \exp{\left(\frac{(1+\epsilon)}{3c_2 r_0^2}-\frac12 \right)}.
			\label{flare-eos1}
		\end{eqnarray}	\\
		\hline
		~~The validity of the wormhole configuration and constraint on the throat radius	& 
		The wormhole configuration is infinitely extended as $ r_0 \le r < \infty $, according to Eq. \eqref{flare-eos1}, with
		\begin{equation}
			r_0^2>\frac{2(1+\epsilon)}{3 c_2}.
			\label{sep-cons_eos1}
		\end{equation}	\\
		\hline
		\hline
		~~\textbf{Case II}			&	\textbf{$ \omega=\frac13 $} \\
		
		\hline
		~~Flaring-out relation	& 
		\begin{eqnarray}
			r < r_0 \left[2 \left(\frac{2(1+\epsilon)}{c_2 r_0^2}-1\right)\right]^{1/6}.
			\label{flare-eos2}
		\end{eqnarray}	\\
		\hline
		~~The validity of the wormhole configuration and constraint on the throat radius	& 
		The wormhole configuration is finitely extended as
		\begin{eqnarray}
			r_0 \le r < r_0 \left[2 \left(\frac{2(1+\epsilon)}{c_2 r_0^2}-1\right)\right]^{1/6}, \qquad \text{with} \qquad r_0^2<\frac{4(1+\epsilon)}{3 c_2}.
			\label{sep-cons_eos2}
		\end{eqnarray}	\\
		\hline
		\hline
		~~\textbf{Case III}			&	\textbf{$ \omega=-1 $} \\
		
		\hline
		~~Flaring-out relation and the validity of the wormhole configuration	& The flare-out condition is satisfied for all $ r\ge r_0 $. Therefore, the wormhole configuration is infinitely extended as $ r_0 \le r < \infty $.	\\
		\hline
		
	\end{tabular}
	\caption{Restrictions imposed by the flaring-out condition for different state parameters in barotropic fluid.}
	\label{details-barotropic_table}
\end{table*}

\subsection{Anisotropic pressure: $ p_t=\sigma p_r,~\sigma \ne 1 $}

If the fluid under consideration is anisotropic in nature then eliminating $ p_r $ and $ p_t $ from equations \eqref{field-eq2} and \eqref{field-eq3} one has the differential equation:
\begin{eqnarray}
	\frac{-r b'(r)+ (1+2\sigma) b(r)}{2 r^3} -(1-\sigma)\left(2a(t) \ddot{a}(t)+\dot{a}(t)^2 - \Lambda a(t)^2 \right) +\frac{\sigma \epsilon}{r^2} -(1-2\sigma) \frac{\gamma a(t)}{r}=0.
	\label{aniso-master}
\end{eqnarray}
As before, for separability of the shape function $ b(r) $ and the scale factor $ a(t) $, one has to choose $ \gamma=0 $ and the resulting differential equations for `$ b $' and `$ a $' take the form:
\begin{eqnarray}
	\frac{-r b'(r)+ (1+2\sigma) b(r)}{2 r^3} +\frac{
	\sigma \epsilon}{r^2} &=& c_3, 
	\label{aniso-separation1} \\
	2a(t) \ddot{a}(t)+\dot{a}(t)^2 - \Lambda a(t)^2 &=& \bar{c}_3,
	\label{aniso-separation2}
\end{eqnarray}
with $ c_3 $ the constant of separation, and $ \bar{c}_3=\frac{c_3}{1-\sigma} $. 

Now solution of \eqref{aniso-separation1} with the throat condition $ b(r_0)=r_0 $ gives the shape function as
\begin{eqnarray}
	b(r) = r \left(\frac{r}{r_0}\right)^{2\sigma} -\epsilon \left[r-r_0\left(\frac{r}{r_0}\right)^{1+2\sigma}\right] - \bar{c}_3 \left[r^3-r_0^3 \left(\frac{r}{r_0}\right)^{1+2\sigma} \right].
	\label{aniso-shape}
\end{eqnarray}

\begin{figure*}[h!]
	\centering
	\includegraphics[height=5cm]{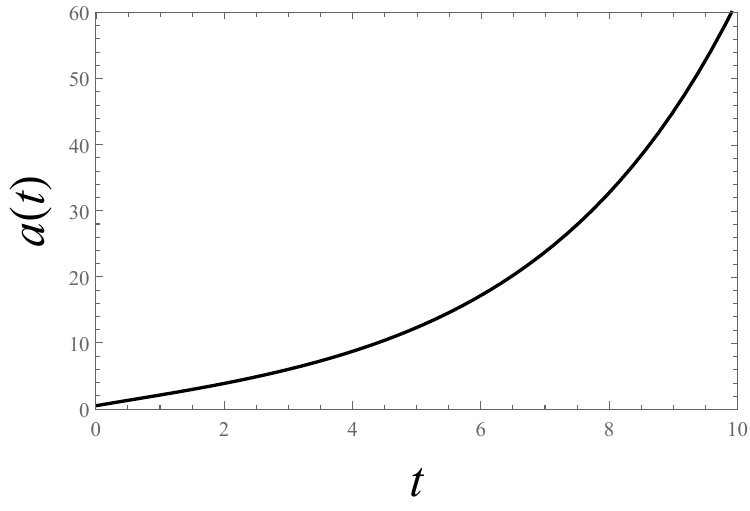}
	\caption{Plots showing the behaviour of scale factor $ a(t) $ of Eq. \eqref{aniso-scale} with cosmic time $ t $ for $\sigma=\frac12,~\Lambda=0.3,~c_3=1$, and $ a_0=0.5 $.}
	\label{fig:aniso-scale1}
\end{figure*}

Similarly, the solution of \eqref{aniso-separation2} for the scale factor can be expressed in an elliptic integral form as
\begin{eqnarray}
	(t-t_0)= \int \frac{da}{\sqrt{\frac{\Lambda}{3}a^2+\frac{A_3}{a}+\bar{c}_3}},
	\label{aniso-scale}
\end{eqnarray}
and with the help of a numerical method, we have obtained the behaviour of scale factor against cosmic time $ t $ as shown in Fig. \ref{fig:aniso-scale1}.

Now, for the validity of flaring-out condition in the wormhole model, it is found that the radial coordinate `$ r $' is restricted as
\begin{eqnarray}
	r < r_0 \left[\sigma \left(1-\frac{(1+\epsilon)}{\bar{c}_3 r_0^2}\right)\right]^{1/2(1-\sigma)} =r_2~\text{(say)}.
	\label{flare-aniso}
\end{eqnarray}
Thus, for the present wormhole configuration, the above expression within the square bracket must be greater than unity, and this restricts the throat radius as
\\ \\
(i) $ 1-\frac{(1+\epsilon)}{\bar{c}_3 r_0^2}>0 $ \qquad i.e. $ r_0^2 > \frac{1+\epsilon}{\bar{c}_3} $,
\\
(ii) $ \sigma \left(1-\frac{(1+\epsilon)}{\bar{c}_3 r_0^2}\right)>1 $ \qquad i.e. $ r_0^2 > - \frac{(1+\epsilon)\sigma}{c_3} $.
\\ \\
Therefore, for the present case, the wormhole geometry is finitely extended with the above restriction on the throat radius for the given restrictions on the constants.

\begin{figure*}[h!]
	\centering
	\subfloat[Traceless fluid with $ c_1=0.0001 $ \label{shape-a}]{{\includegraphics[height=4.5cm]{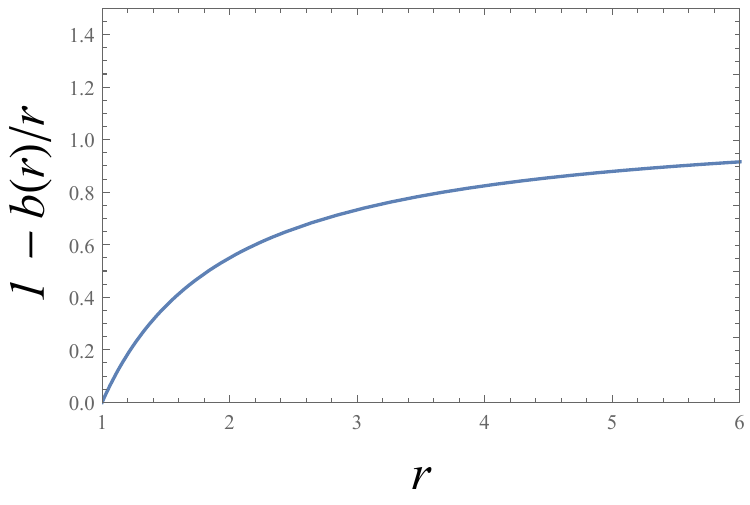}}}\qquad
	\subfloat[Barotropic fluid
	\label{shape-b}]{{\includegraphics[height=4.5cm]{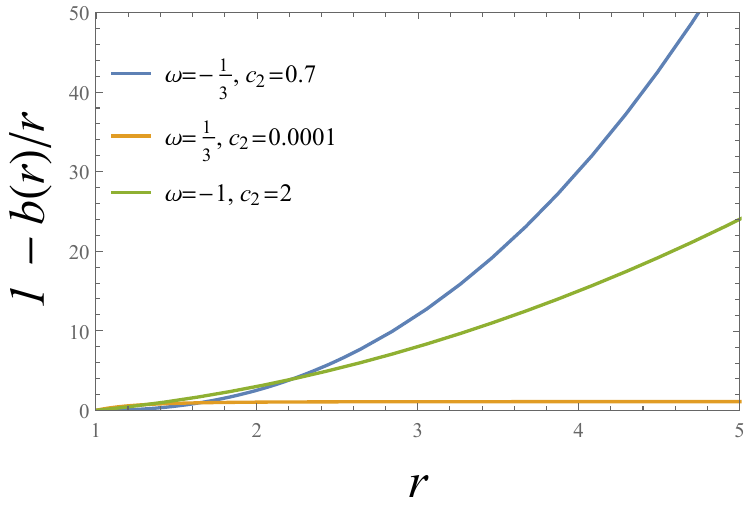}}}\\
	\subfloat[Anisotropic pressure fluid
	\label{shape-c}]{{\includegraphics[height=4.5cm]{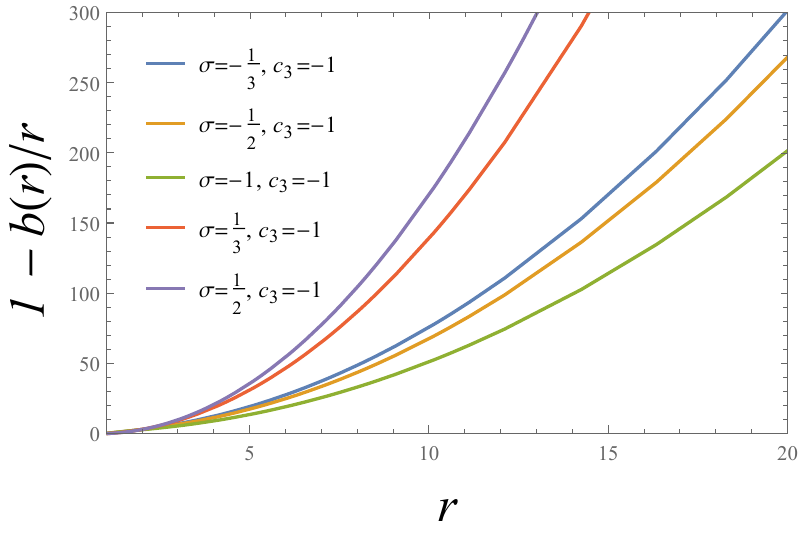}}}
	\caption{Plots showing the behaviour of $ \left(1-\frac{b(r)}{r}\right) $ against the radius $ r $ for three kinds of solutions. The legends in (b) and (c) denote the choice of constants and anisotropy parameters. The throat radius is fixed at $ r_0=1 $ in each plot.}
	\label{shape_plot}
\end{figure*}

Lastly, Fig. \ref{shape_plot} presents a graphical representation of \( \left(1-\frac{b(r)}{r} \right) \) plotted against the radial coordinate \( r \) for the three aforementioned types of wormholes. This figure demonstrates that, with suitable parameter choices, the wormhole configurations generally align with the hyperbolic FRW universe at large radial values. Therefore, these evolving wormhole solutions are consistent with the accelerated expansion, where both universes on either side of the throat are experiencing simultaneous acceleration. This observation can be corroborated in traceless and barotropic fluid systems by the exponential and hyperbolic functions in Eq. \eqref{traceless-scale} and Eq. \eqref{eos-scale2}, respectively.

It is important to note, however, that in the barotropic fluid with \( (\omega=1/3, c_2=0.0001) \), the solution does not align with the hyperbolic FRW universe. Furthermore, a significant dependence on the sign and values of the separation constants is observed, underscoring the additional importance of these parameters.

\section{Energy Conditions}\label{energy-condition}
One of the most interesting and somewhat unique features of traversable wormhole formation is the requirement of exotic matter that violates the classical energy conditions. It is necessary for sustaining the wormhole throat and thereby ensuring traversability. As mentioned in the introduction, this feature has been extensively investigated earlier in evolving traversable wormholes as well as in Morris-Thorne type wormholes in many different versions of modified theories and interestingly some of them support matters that satisfy the energy conditions in various aspects. In particular, even if the standard matter satisfies the conditions, there must be some coupled matter source present that, in turn, acts as exotic energy to fulfill the requirement of traversability. In \cite{epl_paper}, it has been examined in detail in terms of geometrical matter.

Subsequently, the dRGT massive gravity is another major candidate where the presence of non-exotic matter at the throat is intuitively investigated in \cite{Dutta:2023wfg}. It is found that, there is a large possibility where massive gravitons play the role of exotic dark energy and thus, the coupled matter such as the perfect fluid in Einstein gravity satisfies the NEC (null energy condition), WEC (weak energy condition), SEC (strong energy condition), and DEC (dominant energy condition). From the definitions, these conditions can be summarized in principle pressure forms as
\\
(i) NEC : $ \rho+p_r\ge0, ~~ \rho+p_t\ge0 $;\\
(ii) WEC : $ \rho\ge0, ~~ \rho+p_r\ge0, ~~ \rho+p_t\ge0 $;\\
(iii) SEC : $ \rho+p_r\ge 0, ~~ \rho+p_t\ge0, ~~ \rho+p_r+2p_t \ge 0 $;\\
(iv) DEC : $ \rho\ge0, ~~ \rho-|p_r|\ge0, ~~ \rho-|p_t|\ge0 $.

Now, in this section, we are about to check the energy conditions for evolving wormhole solutions in Einstein-massive gravity. The traceless, barotropic and anisotropic fluids have been discussed separately, as follows:
\\ \\
$\bullet$ \textbf{Traceless fluid:}

\begin{figure*}[h!]
	\centering
	\subfloat[$ \rho(r,t) $ \label{en-trace_1}]{{\includegraphics[width=8.2cm]{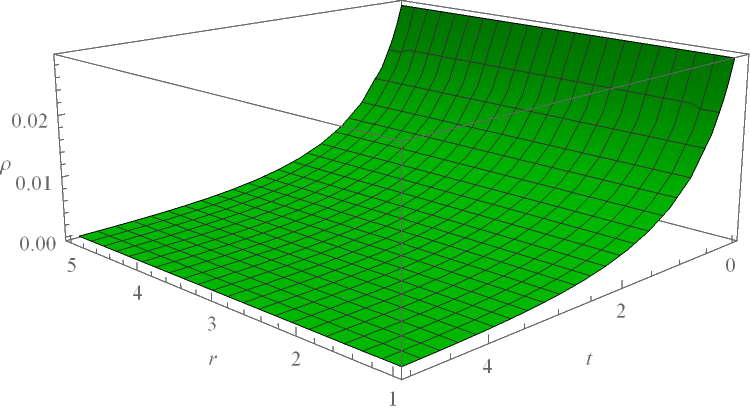}}}\qquad
	\subfloat[$ (\rho(r,t)+p_r(r,t)) $
	\label{en-trace_2}]{{\includegraphics[width=8.2cm]{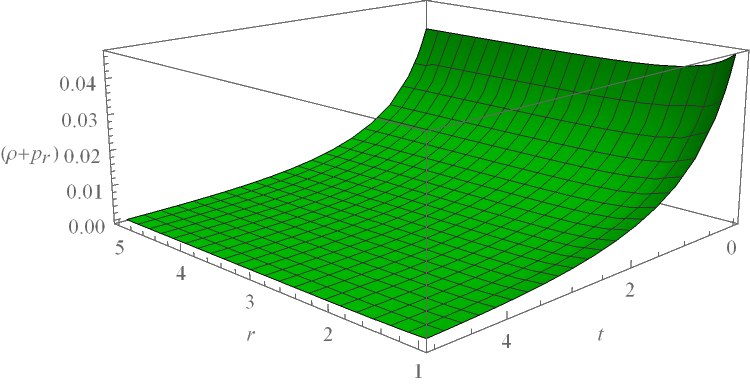}}}
	\\
	\subfloat[$ (\rho(r,t)+p_t(r,t)) $
	\label{en-trace_3}]{{\includegraphics[width=8.2cm]{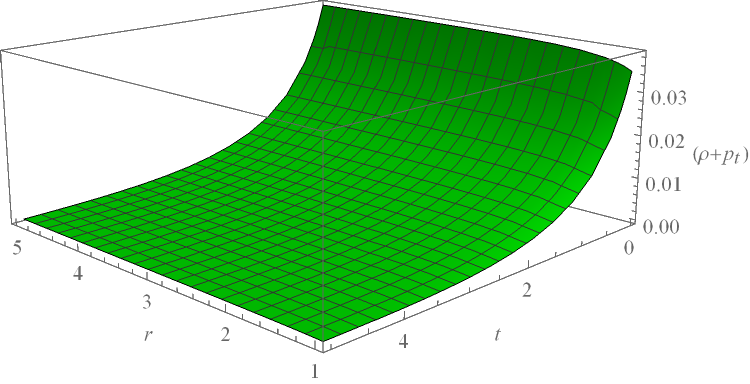}}}\qquad
	\subfloat[$ (\rho(r,t)-|p_r(r,t)|) $
	\label{en-trace_4}]{{\includegraphics[width=8.2cm]{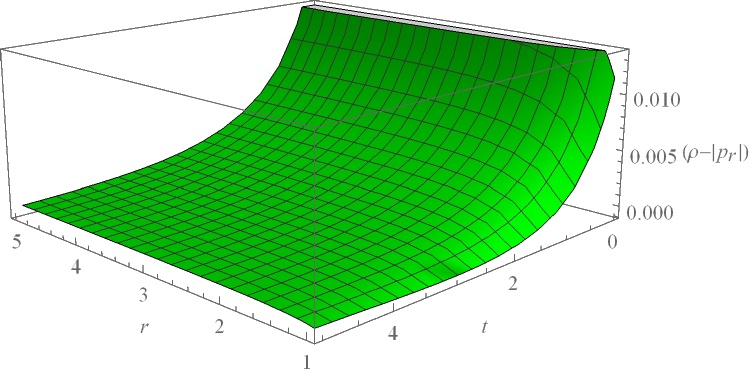}}}
	\\
	\subfloat[$ (\rho(r,t)-|p_t(r,t)|) $
	\label{en-trace_5}]{{\includegraphics[width=8.2cm]{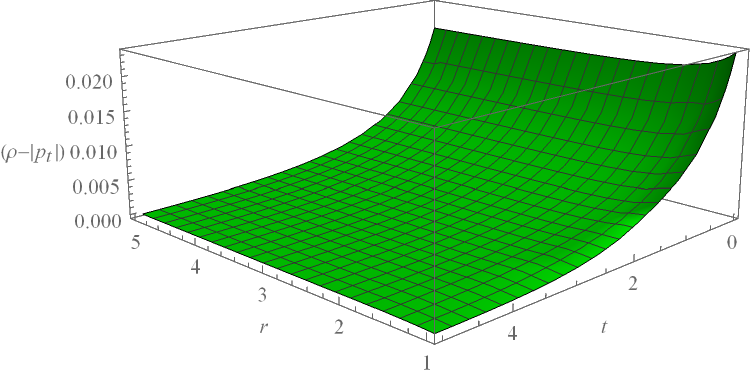}}}\qquad
	\subfloat[$ (\rho(r,t)+p_r(r,t)+2p_t(r,t)) $
	\label{en-trace_6}]{{\includegraphics[width=8.4cm]{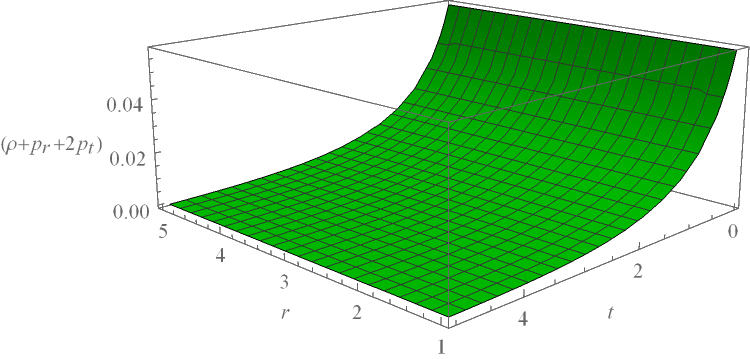}}}
	\caption{Behaviour of energy condition components against radial parameter $ r $ and cosmic time $ t $ for traceless fluid solution. The parameters are set to $r_0=1,~\Lambda=0.3,~c_1=1,~A_1=1,~B_1=-1 $ and $ \epsilon=-1 $ in each plot.}
	\label{en-traceless_plot}
\end{figure*}

For the energy condition components in traceless fluid solution, one may substitute Eq. \eqref{traceless-shape}, \eqref{traceless-scale} into \eqref{field-eq1}, \eqref{field-eq2} and \eqref{field-eq3} to obtain $ \rho(r,t),~(\rho(r,t)+p_r(r,t)),~(\rho(r,t)+p_t(r,t)),~(\rho(r,t)-|p_r(r,t)|),~(\rho(r,t)-|p_t(r,t)|) $ and $ (\rho(r,t)+p_r(r,t)+2p_t(r,t)) $. Therefore, an extensive analysis of the energy conditions are performed for this model for fixed values of throat radius $ r_0=1 $, effective cosmological constant $ \Lambda=0.3 $ and $ c_1=1 $. We have made calculations for different variations of three parameters, namely $ \epsilon,~A_1 $ and $ B_1 $, for which the observations are summarized in Table \ref{en_traceless_table1}. It can be noted that the energy conditions are highly dependent on $ A_1 $ and $ B_1 $. For $ A_1=-1, B_1=1 $, the components contain singularity at a specific instant of time. However, for $ \epsilon=-1,~A_1=1,B_1=-1 $, the energy conditions are completely satisfied for the whole range of $ r $ and $ t $ which can be visualized in a 3D plot as exhibited in Fig. \ref{en-traceless_plot}.

\begin{table*}[h!]
	\centering
	\begin{tabular}{ p {2 cm} p {4 cm} p {9 cm} }
		\hline
		\hline
		~~\textbf{$ \epsilon $}	& \textbf{$ A_1 $ and $ B_1 $}	&	\textbf{Energy condition} \\
		\hline
		~~$ \epsilon=+1 $	& $ A_1=1,~B_1=1 $		& completely violated.	 \\
							& $ A_1=1,~B_1=-1 $		& violated for a certain range in $ t $.	\\
							& $ A_1=-1,~B_1=1 $		& satisfied for a certain time interval with the presence of singularity.	\\
							& $ A_1=-1,~B_1=-1 $	& completely violated.	\\
		\hline
		~~$ \epsilon=-1 $	& $ A_1=1,~B_1=1 $		& completely violated.	 \\
							& $ A_1=1,~B_1=-1 $		& completely satisfied for whole range.	\\
							& $ A_1=-1,~B_1=1 $		& completely satisfied with the presence of singularity.	\\
							& $ A_1=-1,~B_1=-1 $	& completely violated.	\\
		\hline
		\hline
	\end{tabular}
	\caption{Results for the energy conditions in traceless fluid solution for fixed values of throat radius $ r_0=1 $, effective cosmological constant $ \Lambda=0.3 $ and $ c_1=1 $.}
	\label{en_traceless_table1}
\end{table*}

$\bullet$ \textbf{Barotropic fluid with $ \omega=\frac13 $:}

\begin{figure*}[h!]
	\centering
	\subfloat[$ \rho(r,t) $ \label{en-con_a}]{{\includegraphics[width=8.2cm]{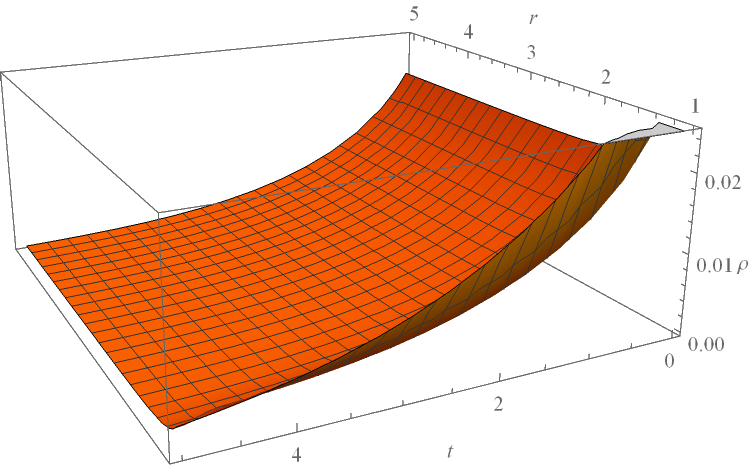}}}\qquad
	\subfloat[$ (\rho(r,t)+p_r(r,t)) $
	\label{en-con_b}]{{\includegraphics[width=8.2cm]{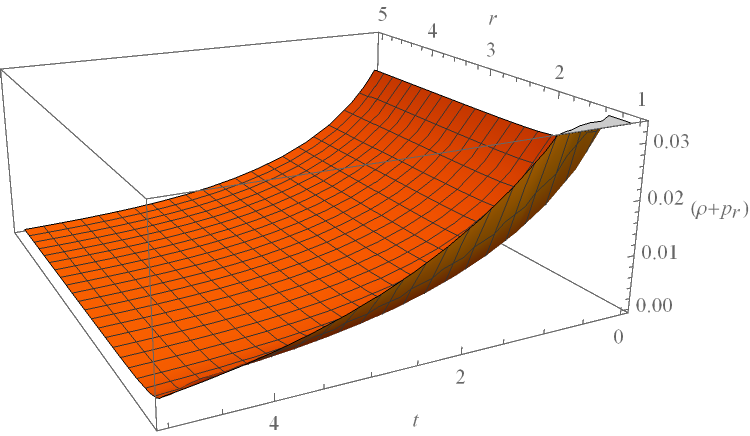}}}
	\\
	\subfloat[$ (\rho(r,t)+p_t(r,t)) $
	\label{en-con_c}]{{\includegraphics[width=8.2cm]{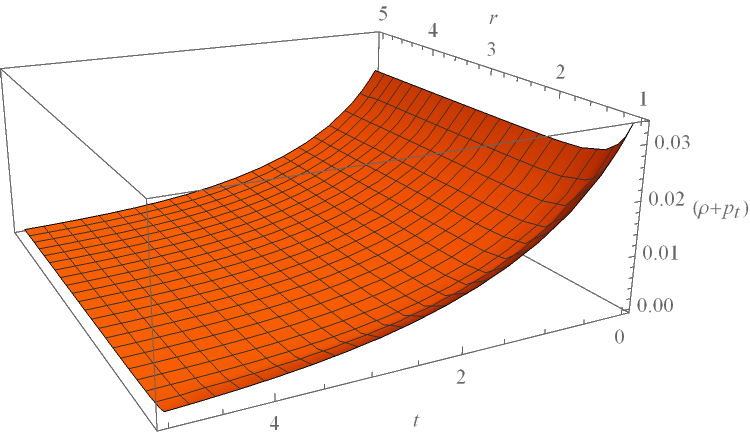}}}\qquad
	\subfloat[$ (\rho(r,t)-|p_r(r,t)|) $
	\label{en-con_d}]{{\includegraphics[width=8.2cm]{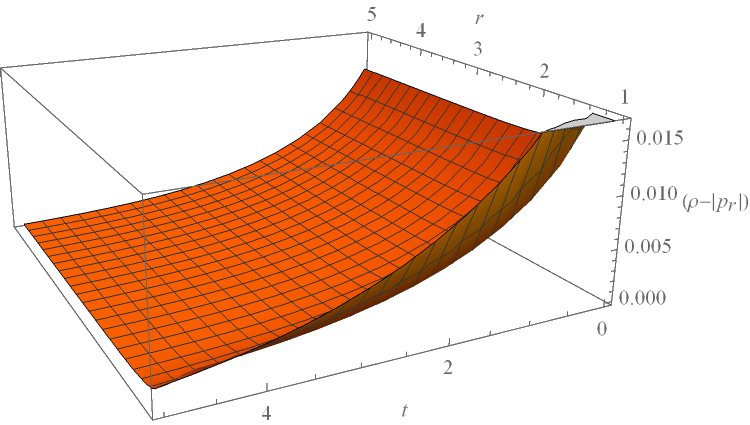}}}
	\\
	\subfloat[$ (\rho(r,t)-|p_t(r,t)|) $
	\label{en-con_c}]{{\includegraphics[width=8.2cm]{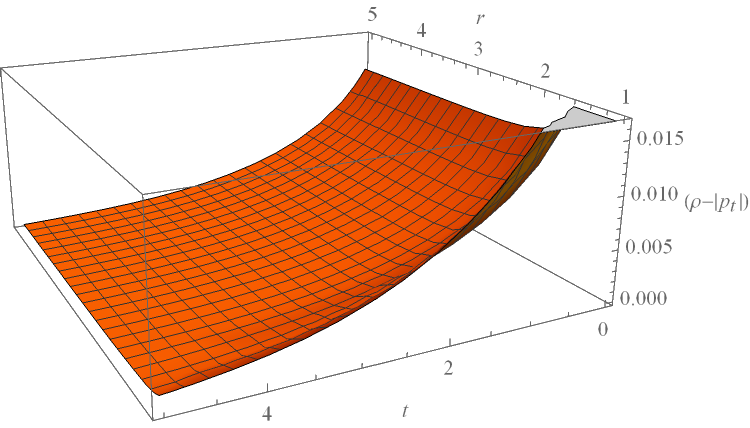}}}\qquad
	\subfloat[$ (\rho(r,t)+p_r(r,t)+2p_t(r,t)) $
	\label{en-con_d}]{{\includegraphics[width=8.4cm]{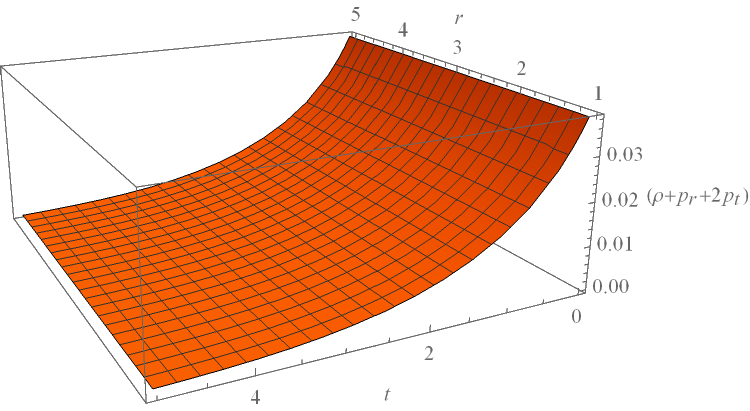}}}
	\caption{Behaviour of energy condition components against radial parameter $ r $ and cosmic time $ t $ for barotropic fluid solution. The parameters are set to $r_0=1,~\Lambda=0.3,~\omega=1/3,~t_0=0,~A_2=1,~c_2=1 $ and $ \epsilon=-1 $ in each plot.}
	\label{en-eos_plot}
\end{figure*}

Now, consider the case of $ \omega=1/3 $ in barotropic fluid to analyze the corresponding energy conditions. Similar to the previous case, taking into account the shape function \eqref{eos-shape}, and scale factor \eqref{eos-scale2}, we substitute them in Eq. \eqref{field-eq1}, \eqref{field-eq2} and \eqref{field-eq3} to obtain the energy condition components $ \rho(r,t),~(\rho(r,t)+p_r(r,t)),~(\rho(r,t)+p_t(r,t)),~(\rho(r,t)-|p_r(r,t)|),~(\rho(r,t)-|p_t(r,t)|) $ and $ (\rho(r,t)+p_r(r,t)+2p_t(r,t)) $. The detailed analysis of the energy condition is investigated by varying $ c_2,~A_2 $ and $ \epsilon $ parameters with a fixed set of values, $ r_0=1,~\Lambda=0.3,~t_0=0 $ and $ \omega=1/3 $. The results are shown in Table \ref{en_eos_table2} where it is observed that $ A_2=1,~c_2=-1 $ imposes singularity on the energy condition components, whereas for $ A_2 $ being negative, the components contain complex values violating the energy conditions. However, for $ \epsilon=-1,~A_2=1,~c_2=1 $, all of them are completely satisfied. This scenario is exhibited in a 3D plot in Fig. \ref{en-eos_plot}. 

\begin{table*}[h!]
	\centering
	\begin{tabular}{ p {2 cm} p {4 cm} p {9 cm} }
		\hline
		\hline
		~~\textbf{$ \epsilon $}	& \textbf{$ A_2 $ and $ c_2 $}	&	\textbf{Energy condition} \\
		\hline
		~~$ \epsilon=+1 $	& $ A_2=1,~c_2=1 $		& completely violated at the throat.	 \\
		& $ A_2=1,~c_2=-1 $		& satisfied for a certain time interval with the presence of singularity.	\\
		& $ A_2=-1,~c_2=1 $		& completely violated.	\\
		& $ A_2=-1,~c_2=-1 $	& completely violated.	\\
		\hline
		~~$ \epsilon=-1 $	& $ A_2=1,~c_2=1 $		& completely satisfied.	 \\
		& $ A_2=1,~c_2=-1 $		& only DEC is violated for a small time interval with the presence of singularity.	\\
		& $ A_2=-1,~c_2=1 $		& completely violated.	\\
		& $ A_2=-1,~c_2=-1 $	& completely violated.	\\
		\hline
		\hline
	\end{tabular}
	\caption{Results for the energy conditions in barotropic fluid solution for fixed values of throat radius $ r_0=1 $, effective cosmological constant $ \Lambda=0.3 $, state parameter $ \omega=1/3 $ and $ t_0=0 $.}
	\label{en_eos_table2}
\end{table*}

$\bullet$ \textbf{Anisotropic fluid:}

Similar to the previous two solutions, energy conditions for the anisotropic fluid is extensively analyzed for fixed values of throat radius $ r_0=1 $, and cosmological constant $ \Lambda=0.3 $. Now, the variation in $ \epsilon,~\sigma $ and $ c_3 $ values shows a decent deviation in the energy condition components, as listed in Table \ref{en_aniso_table}. It is observed that irrespective of the sign in $ \epsilon $, $ \sigma=1/3,~c_3=1 $ shows violation at the wormhole throat, whereas for $ \sigma=-1/3,~c_3=1 $, only DEC is slightly violated at the throat. Apart from that, when $ c_3 $ is negative (i.e., $ c_3=-1 $ with $ \sigma=\pm 1/3 $), all the energy conditions are completely satisfied throughout the spacetime.

\begin{table*}[h]
	\centering
	\begin{tabular}{ p {2 cm} p {4 cm} p {9 cm} }
		\hline
		\hline
		~~\textbf{$ \epsilon $}	& \textbf{$ \sigma $ and $ c_3 $}	&	\textbf{Energy condition} \\
		\hline
		~~$ \epsilon=+0.1 $	& $ \sigma=1/3,~c_3=1 $		& only violated at the throat.	 \\
							& $ \sigma=1/3,~c_3=-1 $	& completely satisfied.	\\
							& $ \sigma=-1/3,~c_3=1 $	& only DEC is violated for a small time interval at the throat.	\\
							& $ \sigma=-1/3,~c_3=-1 $	& completely satisfied.	\\
		\hline
		~~$ \epsilon=-0.1 $	& $ \sigma=1/3,~c_3=1 $		& only violated at the throat.	 \\
							& $ \sigma=1/3,~c_3=-1 $	& completely satisfied.	\\
							& $ \sigma=-1/3,~c_3=1 $	& only DEC is violated for a small time interval at the throat.	\\
							& $ \sigma=-1/3,~c_3=-1 $	& completely satisfied.	\\
		\hline
		\hline
	\end{tabular}
	\caption{Results for the energy conditions in anisotropic fluid solution for fixed values of throat radius $ r_0=1 $ and effective cosmological constant $ \Lambda=0.3 $.}
	\label{en_aniso_table}
\end{table*}

From the above discussions it can be concluded that there is wide range of possibilities for non-exotic matter evolving wormholes in dRGT massive gravity theory. It is however already explained in \cite{Dutta:2023wfg} that for the construction of traversable wormholes in massive gravity theory, massive gravitons play the role of anisotropic dark energy which reflects the violation of energy conditions allowing the coupled matter to be ordinary \cite{epl_paper}. Therefore, for some stringent constraint on the parameter choices, this coupled matter can obey all the classical energy conditions and the wormholes can be constructed with ordinary matter at the throat.

\section{Discussions}\label{discussions}
In this study, we have adopted the technique of smoothly merging spherically inhomogeneous wormhole metrics in the cosmological background of the dRGT massive gravity which is coupled to the Einstein gravity. The model is generated by two fluid systems: one is the perfect fluid supported by GR and the other is massive gravitons having spacial anisotropic dark energy nature embedded within itself. A remarkable feature of the model is the presence of non-zero effective cosmological constant ($\Lambda$) and global monopole potential ($ \epsilon $) generated by the massive graviton mass. The noteworthy aspect is that the energy density responsible for threading and maintaining a wormhole is characterized by its inhomogeneity and anisotropy. This energy-matter component could potentially be a traceless fluid, an ideal barotropic fluid, or any anisotropic cosmic fluid that meets the criteria of homogeneity and anisotropy.

In all the three systems i.e. traceless, barotropic and anisotropic pressure fluids, it is important that with a simple approximation on the massive gravity parameters, i.e. $\gamma=0$, the wormhole shape function and scale factor can be easily separated. At the same time, the approximation does not significantly affect the other parameters. Recalling Eq. \eqref{massive-parameters}, we have $\alpha=-\frac{1+3\beta}{2}$ for $ \gamma=0 $, and consequently, $ \Lambda=-\frac32 m_g^2(1-\beta) $, and $ \epsilon= -m_g^2 c^2 \left(\frac{1-3\beta}{2}\right) $. Thus, we find a constraint on the massive gravity parameters. Subsequently, it is determined that the variations in the wormhole shape function and the scale factor from general relativity are due to terms involving $ \epsilon $ and $ \Lambda $, respectively, in the extension to massive gravity.

Note that, the evolution of the wormhole models obtained in the study are supported by the accelerated expansion in both sides of the throat, however some of them do not agree with the hyperbolic FRW universe models. The most interesting feature coming out of the traceless fluid solution is that the scale factor evolves from an emergent scenario. So, we may say that the corresponding wormhole configuration has evolved from an emergent universe in the past. This aspect has been in accord with the work in reference \cite{Chakraborty:2017efi}.

In section \ref{solution}, we have obtained wormhole solutions in traceless, barotropic and anisotropic fluids and analysed the validity of the solutions corresponding to respective flaring-out conditions. We observed that some of the wormhole solutions are infinitely extended where some of them are confined in finite regions, and therefore we have computed the corresponding region of validity for those finite wormholes.
For a final remark, we may conclude that some of the finitely extended wormhole solutions do not satisfy the flaring-out condition globally so that the global asymptotic flatness can be disturbed and it may have been the result of strong repulsion produced by massive gravitons. This feature has been reflected in Fig. \ref{shape_plot} for all the three cases. As discussed in \cite{Dutta:2023wfg}, the static traversable wormholes violate the asymptotic condition in dRGT massive gravity due to the repulsive effect of gravity. The effect which is produced by the massive gravitons pushes the curvature so strongly that the flatness is disturbed. The accelerated expansion may also have certain effects generated by this repulsion.

In contrast to these three fluid cases, the isotropic fluid ($ p_t=p_r $) solution does not explicitly depend on the scale parameter $ a(t) $. It may have been inevitable, but the solution also rejects any corresponding dependence on massive gravity in terms of effective cosmological constant. However, if one may proceed with the leftover expressions by considering a reasonable choice of $ a(t) $ with $ \dot{a}(t)/a(t)=H(t) \ne 0 $, by plugging-in the inflation mechanism within the framework, there still have been an infinite set of choices for $ a(t) $. Nevertheless, the possible solution of the shape function in the form $ b(r)=r^3/r_0^2 $, does not relate to well-defined asymptotic structure since the flaring-out condition ($ -r b'(r)+b(r)= -2r^3/r_0^2 $) is \textit{necessarily violated} in all space. Although, the physical picture may still have something to present. If we choose a competent scale factor as $ a(t)=a_0 e^{H_0 t} $, the metric takes the form
\begin{eqnarray}
\nonumber
	ds^2=-dt^2+a_0^2 e^{2H_0 t} \left[\frac{dr^2}{1-\frac{r^2}{r_0^2}}+r^2 d\Omega^2 \right].
\end{eqnarray}
The spatial form of the metric denotes an exponentially expanding 3-sphere, and thereby represents an empty closed universe satisfying $ \frac{1}{r_0^2}>0 $.

One of the most remarkable aspects of massive gravity theory is the behaviour of the massive graviton energy-momentum tensor which naturally fulfills the energy condition violation due to its anisotropic dark energy nature. As mentioned, the static spherically symmetric traversable wormhole in massive gravity satisfies all the energy conditions \cite{Dutta:2023wfg}, so there may be a presence of ordinary matter to thread the wormhole throat. In the context of evolving wormholes, this paper intended to study the interaction of perfect fluid and massive gravitons in traceless, barotropic, and anisotropic fluids. In the context of all three fluid solutions, numerous parameter selections result in wormhole configurations featuring non-exotic matter at the throat. However, it is also possible that for some of these choices, the energy condition components at the throat transition from positive to negative values over cosmological time, indicating the evolution of the matter content from non-exotic to exotic.

For a final conclusion we may say that it is important to observe that the geometries of the evolving wormholes, when situated far from the throat, resemble a flat FRW universe. At first glance, if the throat of the wormhole is positioned beyond the cosmological horizon of any observer, that observer is not causally connected to the throat. Therefore, for extended periods, an observer within this wormhole universe, located sufficiently distant from the throat, will perceive the universe as isotropic and homogeneous. In such a scenario, the observer may find it challenging to distinguish whether they inhabit a space with constant curvature or a space within a wormhole spacetime.

\section*{Acknowledgement}
The authors are extremely thankful to the anonymous referee whose suggestions and comments improved the quality and visibility of the paper. S.C. thanks FIST program of DST, Department of Mathematics, JU (SR/FST/MS-II/2021/101(C)).

\end{document}